\documentclass[12pt,a4paper,final]{article}
\usepackage{graphicx}

\begin{document}

\title{Development of a low-energy radioactive ion beam facility for the MARA separator}
\author{Philippos Papadakis\footnote{philippos.papadakis@jyu.fi}, Iain Moore, Ilkka Pohjalainen, \\Jan Sar\'en and Juha Uusitalo \\ \\ {\small University of Jyv\"askyl\"a, Department of Physics, P.O. Box 35,} \\ {\small FI-40014 University of Jyv\"askyl\"a, Finland.}}
\date{}
\maketitle

\begin{abstract}
A low-energy radioactive ion beam facility for the production and study of nuclei produced close to the proton drip line is under development at the Accelerator Laboratory of the University of Jyv\"askyl\"a, Finland. The facility will take advantage of the mass selectivity of the recently commissioned MARA vacuum-mode mass separator. The ions selected by MARA will be stopped and thermalised in a small-volume gas cell prior to extraction and further mass separation. The gas cell design allows for resonance laser ionisation/spectroscopy both in-gas-cell and in-gas-jet. The facility will include experimental setups allowing ion counting, mass measurement and decay spectroscopy.
\end{abstract}

\section{Introduction}
\label{intro}
The quest to answer fundamental questions on the properties and structure of the elements and their creation through astrophysical processes has led to the study of short-lived exotic nuclei produced far from stability. This became possible due to the development of an array of tools and techniques which allow chemical and/or physical separation of the different elements, or even different isotopes, based on properties such as mass, charge, atomic levels and chemical reactivity with various substances.

The MARA low-energy branch (MARA-LEB), under development at the Accelerator Laboratory of the University of Jyv\"askyl\"a (JYFL-ACCLAB), will employ a combination of techniques for the separation of atoms produced through fusion-evaporation reactions at the MARA (Mass Analysing Recoil Apparatus) separator \cite{mara}. In this paper the current status of the MARA-LEB facility will be discussed in light of recent developments at JYFL-ACCLAB and other facilities.

\section{Scientific motivation}
\label{sec:motivation}
How are the elements created in the cosmos? What underlying microscopic and macroscopic phenomena take place in exotic nuclei? Are there new modes of decay far from stability? These are some of the general scientific questions that have motivated the development of the MARA-LEB facility, which aims to study nuclei produced in weakly-populated fusion-evaporation channels by separating them from other overwhelmingly abundant reaction products. Three cases of particular interest are discussed in the following. 
\begin{itemize}
\item Studies of nuclear structure and properties of $^{80}$Zr.
\end{itemize}
Nuclei with N$\sim$Z in the mass region of A$\sim$80, such as $^{80}$Zr (N=Z=40), can be used as testing grounds for different nuclear models \cite{naz}. In these nuclei the addition or subtraction of a few nucleons leads to dramatic changes in shape as proton and neutron shell effects act coherently giving rise to a coexistence of different nuclear shapes. In the case of $^{80}$Zr theoretical calculations indicate a coexistence of five different nuclear shapes, which if verified experimentally would be a unique example of shape coexistence in the nuclear chart \cite{rodr}. $^{80}$Zr is also of astrophysical interest since it is considered to be a waiting point in the astrophysical rapid proton capture process (rp-process \cite{wallace}) as when undergoing proton capture it leads to the proton-unbound $^{81}$Nb \cite{schatz}. Resonance laser ionisation studies of $^{80}$Zr will allow the direct investigation of deformation and the changes in mean-square charge radii, while mass measurements in the region will provide information on the existence of the ZrNb cycle \cite{schatz2}.

\break
\begin{itemize}
\item Investigation of $^{94}$Ag and its isomers.
\end{itemize}
N=Z $^{94}$Ag is one of the nuclei whose precise mass determination is crucial when calculating the final elemental abundances and the path of the rp-process. This isotope has a (21$^+$) spin trap isomer with the highest spin observed in $\beta$-decaying nuclei. The high excitation energy and spin of the isomer, together with its long half-life, allow for a multitude of decay modes such as $\beta$- and one-proton decay. Evidence of two-proton decay was reported for this isomer and was attributed to strong prolate deformation \cite{mukha}. However large-scale shell model calculations \cite{kaneko} and further experiments \cite{pechenaya}\cite{cerny}\cite{kankainen} failed to confirm the existence of this decay mode bringing to question the previous findings. A detailed study of this nucleus through direct mass measurements and resonance ionisation spectroscopy will provide unambiguous determination of the energy of the high-spin isomer and model-independent information with respect to the postulated strong prolate deformation.

\begin{itemize}
\item Region of the N=Z=50 nucleus $^{100}$Sn.
\end{itemize}
The $^{100}$Sn region is expected to provide a wealth of information on the validity of shell-model predictions and the evolution of nuclear shell gaps away from the valley of stability. However, low production cross sections have hindered such studies. For example, reliable data on ground-state properties is available only until $^{108}$Sn \cite{eberz}. An extensive program to study tin nuclei towards the neutron deficient isotope $^{102}$Sn, through in-gas-jet resonance ionisation spectroscopy, will be feasible at MARA and will provide significant insights on the hyperfine structure and changes in mean square charge radii.

\section{Production of low-energy radioactive ion beams at MARA}
\label{sec:leb}
\begin{figure}
\includegraphics[width=\textwidth]{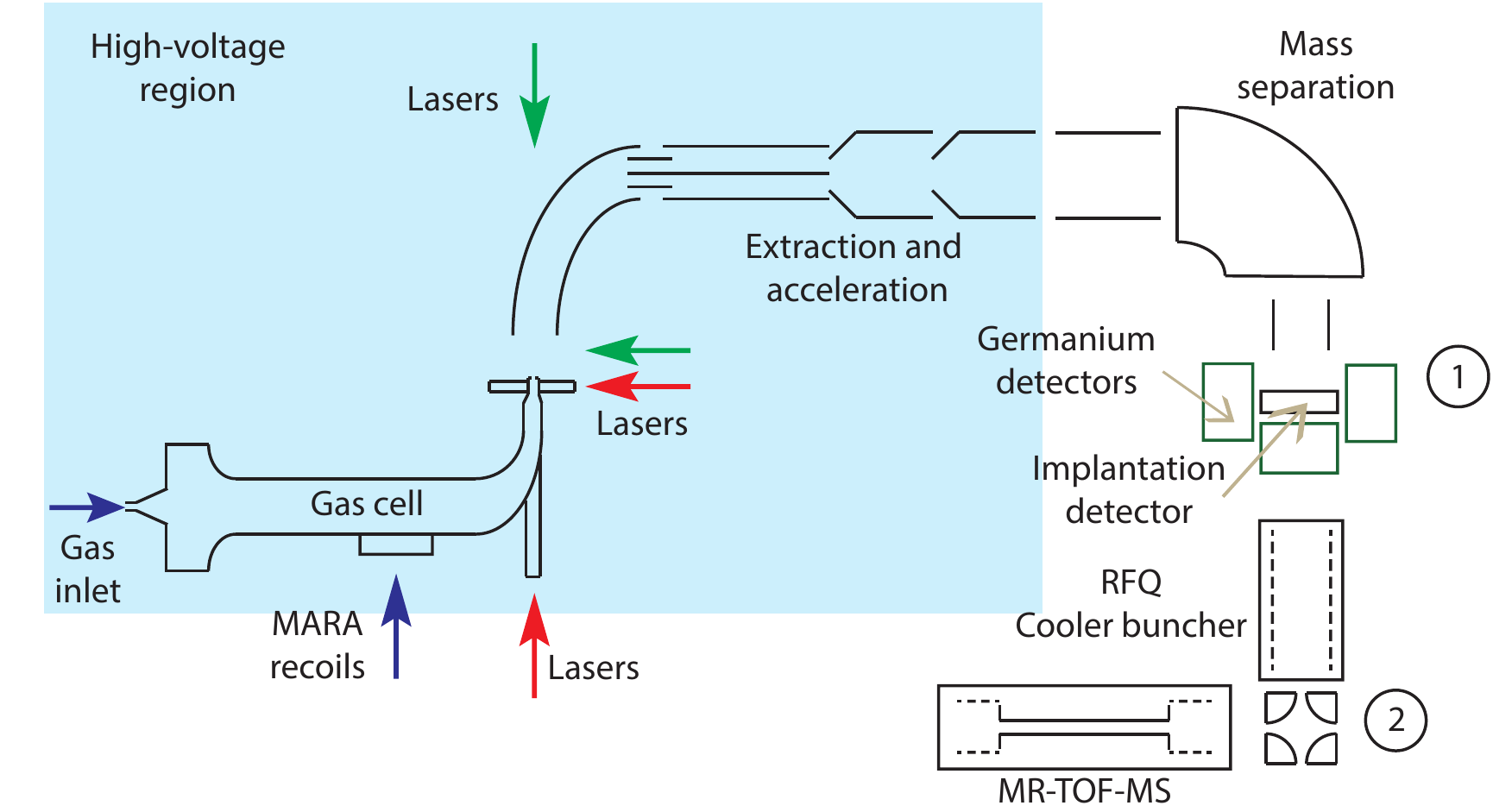}
\caption{A schematic representation of the concept of the MARA-LEB facility. The MARA EVRs will be stopped inside a gas cell, which offers access for laser ionisation from different ports. Lasers indicated with red colour are for in-gas-cell ionisation and the ones with green are for in-gas-jet ionisation. The gas cell and radiofrequency ion guides will be on high voltage while the mass separating dipole magnet and detector stations will be at ground potential. Two experimental systems, a compact decay station (1) and a mass measurement setup (2) are shown in the figure. Parts are not to scale}
\label{fig:LEB-concept}
\end{figure}

Low-energy radioactive ion beams (RIBs) will be produced at MARA through fusion-evaporation reactions employing stable beams and targets. At MARA the evaporation residues (EVRs) will be separated according to their mass and charge states by a combination of static electric and magnetic fields. Nuclei which fulfil the chosen selection criteria will enter a buffer gas cell located at the focal plane, where they will be stopped, thermalised and, depending on the buffer gas used, neutralised. If the atoms are neutralised they will be selectively ionised using multi-step laser resonance ionisation either within an ionisation volume inside the gas cell or at the exit of the cell. After extraction from the cell the ions will be transported by a series of radiofrequency ion guides towards an acceleration stage and then to a dipole magnet for further mass separation. The ions of interest will then be transported towards the experimental stations. The experimental stations to be used with the MARA-LEB facility will include a compact decay station and a multi-reflection time-of-flight mass spectrometer (MR-TOF-MS) \cite{wolf} coupled to a radiofrequency cooler buncher \cite{nieminen} required for its operation. MARA and the MARA-LEB facility are described in the following while a schematic representation of the concept is shown in Fig.~\ref{fig:LEB-concept}.

\subsection{The MARA vacuum-mode recoil separator}
\begin{figure}
\includegraphics[width=\textwidth]{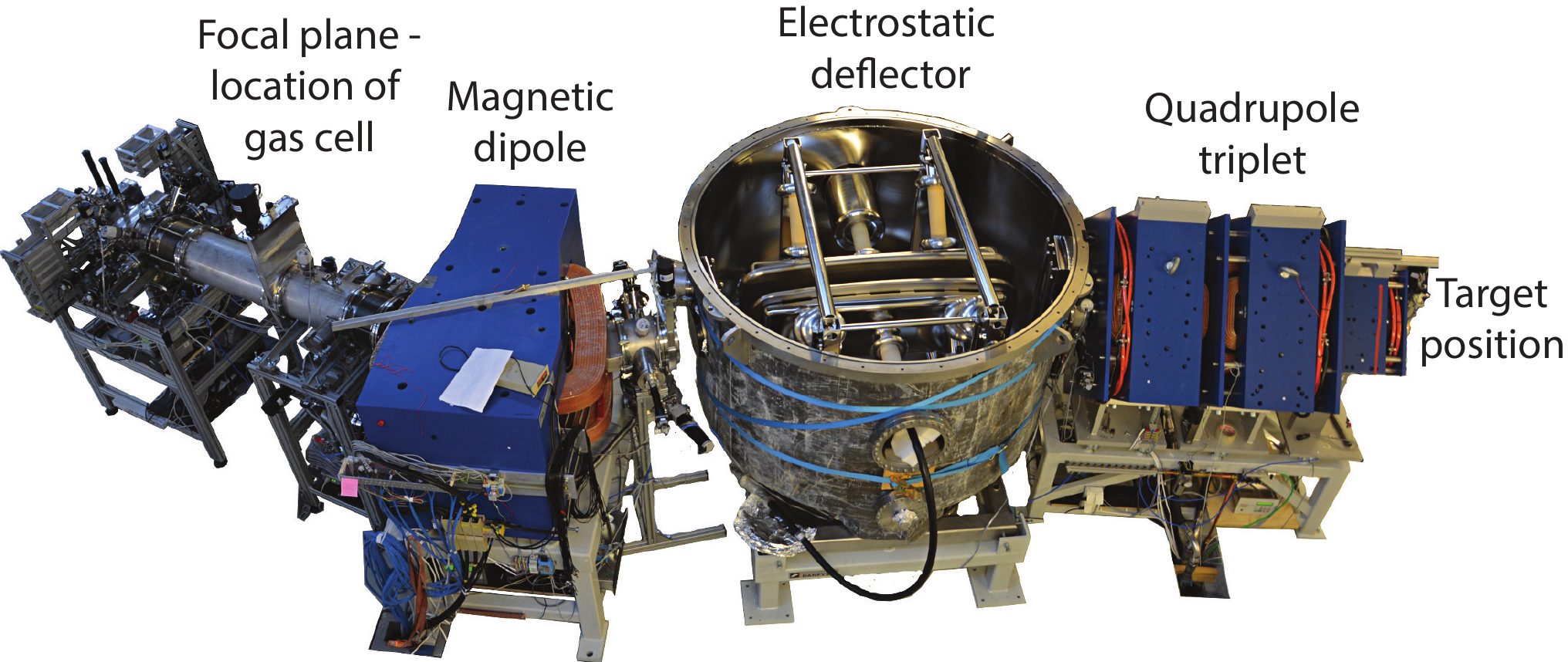}
\caption{A picture of the MARA separator with the electrostatic deflector chamber open. The main parts of the separator are indicated in the figure}
\label{fig:mara}
\end{figure}

The MARA vacuum-mode recoil separator (Fig.~\ref{fig:mara}) consists of a magnetic quadrupole triplet followed by an electrostatic deflector and a magnetic dipole (QQQED configuration). It provides a first order mass resolving power of $\sim$250, nominal dispersion of 8.0\,mm/(\% in m/q) and has an angular acceptance of 10\,msr. A rotating target wheel will allow the use of primary beams with intensities of the order of a few hundred pnA. 

\begin{figure}
\begin{center}
\includegraphics[width=.75\textwidth]{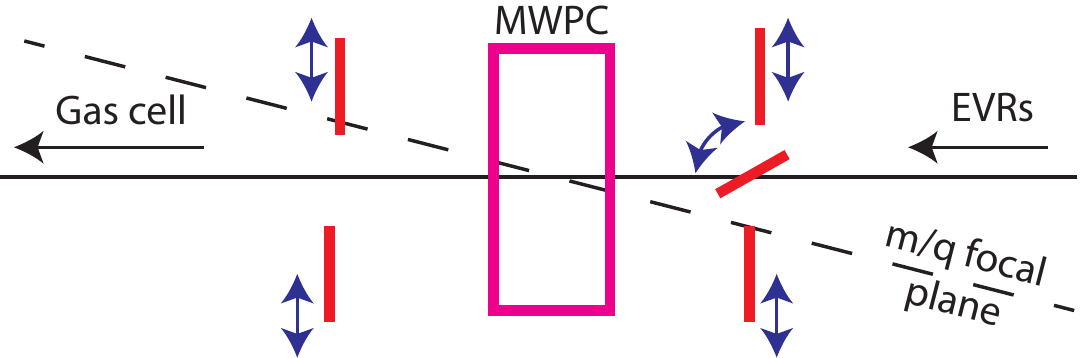}
\caption{A representation of the mechanical slits system of MARA. The slits are shown with thick red lines and the MWPC with a magenta box. The m/q focal plane of MARA is tilted by 14.4$^\circ$. The figure is reproduced from \cite{mara} and is not to scale}
\label{fig:mara-slits}
\end{center}
\end{figure}

In order to suppress ions with the wrong mass or charge from reaching the focal-plane detectors a system of mechanical mass slits is used. A combination of two sets of slits (one before and one after the focal plane) moving perpendicularly to the EVR axis and one rotatable slit was shown to effectively suppress contaminant ions during recent commissioning runs. The rotating slit could be placed in the centre of the perpendicular mass slits located either upstream of the multiwire proportional counter (MWPC) or downstream, or it can be removed from the system if not needed. Two sets of mass slits are required as the m/q focal plane of MARA is tilted by 14.4$^\circ$ (see \cite{mara} for further details). A schematic representation of the slit system is shown in Fig.~\ref{fig:mara-slits}.

The performance of MARA was investigated for symmetric (e.g. $^{40}$Ar+$^{45}$Sc) and asymmetric reactions, both in normal (e.g. $^{40}$Ar+$^{124}$Sn) and inverse (e.g. $^{78}$Kr+$^{58}$Ni) kinematics and was found to be within the design criteria and in agreement with the simulations published in \cite{mara}.

\subsection{Transmission efficiency and image at the focal plane of MARA}
\label{subsec:distribution}
The good agreement between ion-optical simulations and commissioning experiments allows one to estimate different experimental parameters (e.g. transmission efficiency, charge state distribution and spatial distribution of the recoils at the focal plane) using the simulations. These parameters typically depend on the type of reaction, the reaction kinematics and the beam and target properties. For example, the transmission efficiency is reduced when a thicker target is used or when the reaction channel of interest includes the emission of an $\alpha$ particle. Additionally the physical separation of charge states at the focal plane is reduced and their image is better focused when a reaction is studied in inverse instead of normal kinematics. Several reactions have been simulated using a simulation code \cite{mara} developed for MARA. The total transmission efficiency when two charge states are collected was found to range from $\sim$15\% to $\sim$30\% depending on the reaction. 

\begin{figure}
\begin{center}
\includegraphics[width=\textwidth]{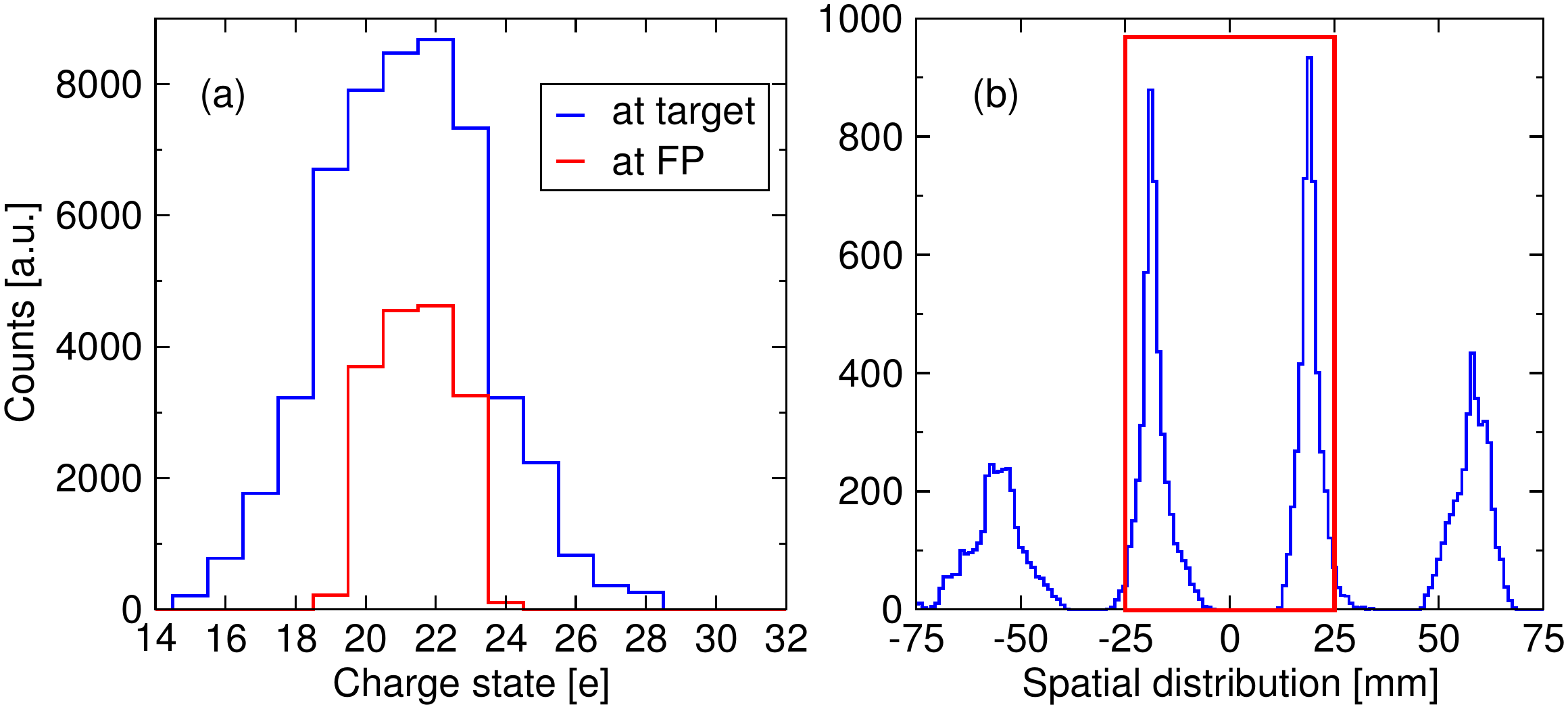}
\caption{(a) Transmission of charge states at the focal plane (FP) for $^{94}$Ag compared with the distribution at the target position. (b) Position spectra of the different charge states at the focal plane. The proposed 50\,mm-wide gas cell window is indicated with a red square}
\label{fig:states}
\end{center}
\end{figure}

One of the most challenging experiments for the MARA-LEB facility will be the investigation of $^{94}$Ag and, as such, it will be used here as an example. Simulations indicate that $\sim$33\% of the $^{94}$Ag nuclei produced at the target position through the $^{58}$Ni($^{40}$Ca,p3n)$^{94}$Ag reaction and when using a 500\,$\mu$g/cm$^2$ target are included within 2 charge states, as shown in Fig.~\ref{fig:states} (a). The transmission efficiency through MARA for these two charge states is calculated as $\sim$53\% leading to the total efficiency of $\sim$18\%. For other reactions relevant to the MARA-LEB facility the transmission efficiency for the two most abundant states could be as high as 80\% giving total efficiencies close to 30\%. Fig.~\ref{fig:states} (b) indicates the position spectra at the focal plane for different charge states. Approximately 99\% of the counts in the two most abundant charge states are included within the 50\,mm-wide red rectangle which represents the possible width of the gas cell entrance window. The surface coils of the MARA dipole can be utilised to adjust the focal length of the separator and move the focus to the gas cell entrance window.

\subsection{A small volume gas cell for stopping EVRs}
\label{subsec:gas_cell}
Buffer gas cells are used widely as a means to stop and thermalise reaction products in order to extract them for further study. Two main methods that use noble gases to stop nuclear reaction products are typically employed: the ion guide technique \cite{igisol}\cite{igisol2} and the ion catcher technique (see for example \cite{catcher}). In the former a laminar gas flow (usually helium) is responsible for the evacuation of the ions from a small-volume gas cell with sufficiently short extraction times in order to minimise the recombination of recoil ions with electrons. The latter method is used mainly with in-flight separators, where a large volume of gas is required in order to contain the full straggling range of the reaction products. In such large volume cells, extraction of ions using gas flow alone is not practical dew to long extraction times and therefore DC electric fields are employed. Additionally, RF structures may be used in the form of carpets or funnels in order to reduce losses caused by ions colliding with the walls \cite{carpets}. 

In order to stop the EVRs from MARA a small-volume gas cell based on the design developed for the REGLIS facility at S3, GANIL (France) \cite{s3-reglis} is under development. The MARA EVRs will enter the gas cell through an entrance window which will separate the high-vacuum region of MARA from the buffer gas volume. Different materials, such as Mylar and Havar, are under consideration for the entrance window. The window thickness will be chosen such that the energy of the EVRs will degrade sufficiently prior to entering the gas so that they are stopped in the centre of the gas cell in order to increase the extraction efficiency \cite{s3-reglis}. As an example, simulated $^{94}$Ag ions produced through the reaction of $^{40}$Ca at 192\,MeV energy on a $^{58}$Ni target of 500\,$\mu$g/cm$^2$ were used as input to SRIM \cite{srim} to calculate the required window thicknesses. 11\,$\mu$m of Mylar would degrade the energy of the ions sufficiently for them to be stopped in $\sim$13\,mm of argon at 500\,mbar pressure with straggling of $\sim$3\,mm. Similarly, 1\,$\mu$m of Mylar combined with 3\,$\mu$m of Havar would require $\sim$17\,mm of argon at 500\,mbar pressure to stop the ions, with similar straggling. As such a 30\,mm-deep cell would be sufficient to stop the $^{94}$Ag ions. In both cases the windows of the MARA MWPC (total of 1\,$\mu$m of Mylar), which will be positioned in-front of the gas cell and will be used for adjusting the settings of MARA for the EVRs of interest, are included in the quoted thicknesses. The possibility of modifying the MWPC so it can be removed after adjustment of the MARA settings without disturbing the alignment of the system is under consideration. This could increase the transmission to the gas cell by varying amounts depending on the experiment. In such a case a 4\,$\mu$m Havar window will allow for similar stopping ranges and straggling as is achieved with the 11\,$\mu$m Mylar window.

The gas cell will be used with argon buffer gas for efficient neutralisation of the ions and subsequent in-gas-cell or in-gas-jet laser ionisation/spectroscopy, or helium buffer gas, when fast extraction times are preferred over neutralisation for the cases where the ions will be transported to the MR-TOF-MS for mass measurements. As such the MARA-LEB vacuum system must be able to sufficiently evacuate both these gases from the ion transport region. A vacuum chamber separated into three individual volumes pumped by either screw or turbo pumps will provide the necessary vacuum conditions. The first volume will house the gas cell and the first ion guide and will be connected to the second chamber through a small size ion guide, which will act as the first differential pumping port. The second chamber will house part of a larger ion guide and will be connected through a differential pumping aperture with the third chamber housing the rest of the ion guide and the acceleration and ground electrodes. The ion optics are under development and will be presented elsewhere. The concept is similar to that presented in \cite{s3-reglis} with the major difference being the use of a 90$^\circ$-bent first ion guide, due to space restrictions, and two, instead of one, electrodes at the extraction point for reaching the desired acceleration voltage. Sufficient inner diameter of the first ion guide will allow for a laser beam with the same diameter as the gas jet to pass through it and overlap with the jet at the exit of the gas cell. The bend removes the requirement for the laser beam to pass through all the ion guides thus allowing for smaller diameter ion guides and hence smaller diameter differential pumping ports. In contrast to the concept presented in \cite{s3-reglis} the gas cell and the MARA-LEB vacuum chamber will be on high-voltage with the ions accelerated towards ground potential.

High purity gas with gas impurities at the order of sub-parts-per-billion are required for the successful operation of the MARA-LEB facility. Impurities of the part-per-million level affect the charge exchange processes and the ion guide operation due to molecular formation. The importance of high gas purity was demonstrated in \cite{kessler} and \cite{moore} especially for chemically active elements. The gas handling system was designed in-house based on that developed for IGISOL-4 \cite{pohjalainen}. It includes liquid nitrogen cold traps for the purification of helium and an active heated getter purifier (PS4-MT3-R-2 SAES MonoTorr) for the purification of argon. Both the gas lines (electropolished stainless steel) and the gas cell will be heated and pumped prior to each experiment for the removal of water vapour from the system.

As shown in Fig.~\ref{fig:LEB-concept} the gas cell is horn shaped allowing for a separate laser ionisation volume, in order to reduce direct recombination of laser ionised atoms in the presence of the weak plasma created by the MARA recoils and the unwanted masses and to allow the use of ion collectors for the collection of non-neutralised ions prior to laser ionisation \cite{dual-cell}. The ions are extracted from the cell solely by the gas flow and no electric fields are used. Laminar gas flow is achieved in the main gas cell volume by using a flow straightening structure at the gas entrance point of the cell as described in \cite{s3-reglis}. The shape of the cell and the flow straightening structure were refined through detailed calculations using the COMSOL package \cite{comsol}. 

Evacuation times for the S3 prototype gas cell were calculated for argon buffer gas using a $^{254}$No simulated beam and are presented in \cite{s3-reglis}. The evacuation time was found to be $\sim$630\,ms for a 30\,mm-deep cell using a 1\,mm exit hole diameter and was reduced to $\sim$400\,ms for a 20\,mm cell. A larger exit hole diameter of 1.5\,mm decreases the evacuation time for the 20\,mm cell to $\sim$190\,ms but will nonetheless increase the gas load on the differential pumping system. Further calculations for helium, performed with a different technique, indicate evacuation times of $\sim$180\,ms for the 30\,mm cell with 1\,mm exit hole diameter, which decreases to $\sim$125\,ms for the 20\,mm cell and the same diameter. The maximum allowed exit hole diameter will be determined by the pumping capacity of the vacuum system. The evacuation time of the gas cell will not have an impact on the gas cell efficiency for the physics cases mentioned in Sec.~\ref{sec:motivation}.

\subsection{Expected yields at the experimental stations}
\label{subsec:yields}
The expected yields at the experimental stations for reactions producing the selected nuclei mentioned in Sec.~\ref{sec:motivation} are presented in Tab.~\ref{tab:nuclei}. These are estimated by assuming primary beam intensities of 200\,pnA, target thicknesses of 500\,$\mu$g/cm$^2$ and a total efficiency for MARA ranging from 15\% to 30\% depending on the reaction. For in-gas-jet resonance ionisation, a gas cell thermalisation, diffusion and transport efficiency to the exit nozzle of 50$\%$ has been assumed, a nominal efficiency of 50$\%$ for neutralisation (unknown and element dependent) and an in-gas-jet laser ionisation efficiency of 50$\%$ (including temporal and spatial overlap) is used. When the ions are transported to the MR-TOF-MS for mass measurements they will not be neutralised prior to their extraction from the gas cell and therefore the neutralisation and in-gas-jet ionisation loss factors are regained. The transport efficiency from the gas cell to the radiofrequency cooler buncher is estimated to be 90$\%$ and the transmission through the buncher to 60$\%$. All of the efficiencies are estimated and must be experimentally determined, however the estimates are conservative. Estimated yields for selected reactions at the REGLIS facility are presented in \cite{ferrer}.

\begin{table}
\small
\caption{Reactions for the production of selected key nuclei and the expected yields at different experimental stations. RIS refers to resonance laser ionisation spectroscopy. All values are approximate. Some of the cross sections have been measured, others are estimated}
\label{tab:nuclei} 
\begin{tabular}{lcccc}
\hline\noalign{\smallskip}
Reaction & Cross & Yield at & Yield after  & Yield at \\
& section & focal plane & in-gas-jet RIS & MR-TOF-MS  \\
& ($\mu$b) & (ions/s) & (ions/s) & (ions/s) \\
\noalign{\smallskip}\hline\noalign{\smallskip}
$^{24}$Mg($^{58}$Ni,2n)$^{80}$Zr & 10 \cite{lister} & 47 & 6 & 13 \\
$^{58}$Ni($^{40}$Ca,p3n)$^{94}$Ag & 0.65 \cite{ferrer} & 1.5 & 0.2 & 0.4 \\
$^{50}$Cr($^{58}$Ni,$\alpha$2n)$^{102}$Sn & 2 \cite{lipoglavsek} & 2.66 & 0.3 & 0.7 \\
$^{92}$Mo($^{16}$O,2n)$^{106}$Sn & 3000 \cite{benjelloun} & 3500 & 440 & 945 \\
\noalign{\smallskip}\hline
\end{tabular}
\end{table}

\subsection{The MARA-LEB facility experimental area}
\label{subsec:area}
\begin{figure}[t!]
\begin{center}
\includegraphics[width=\textwidth]{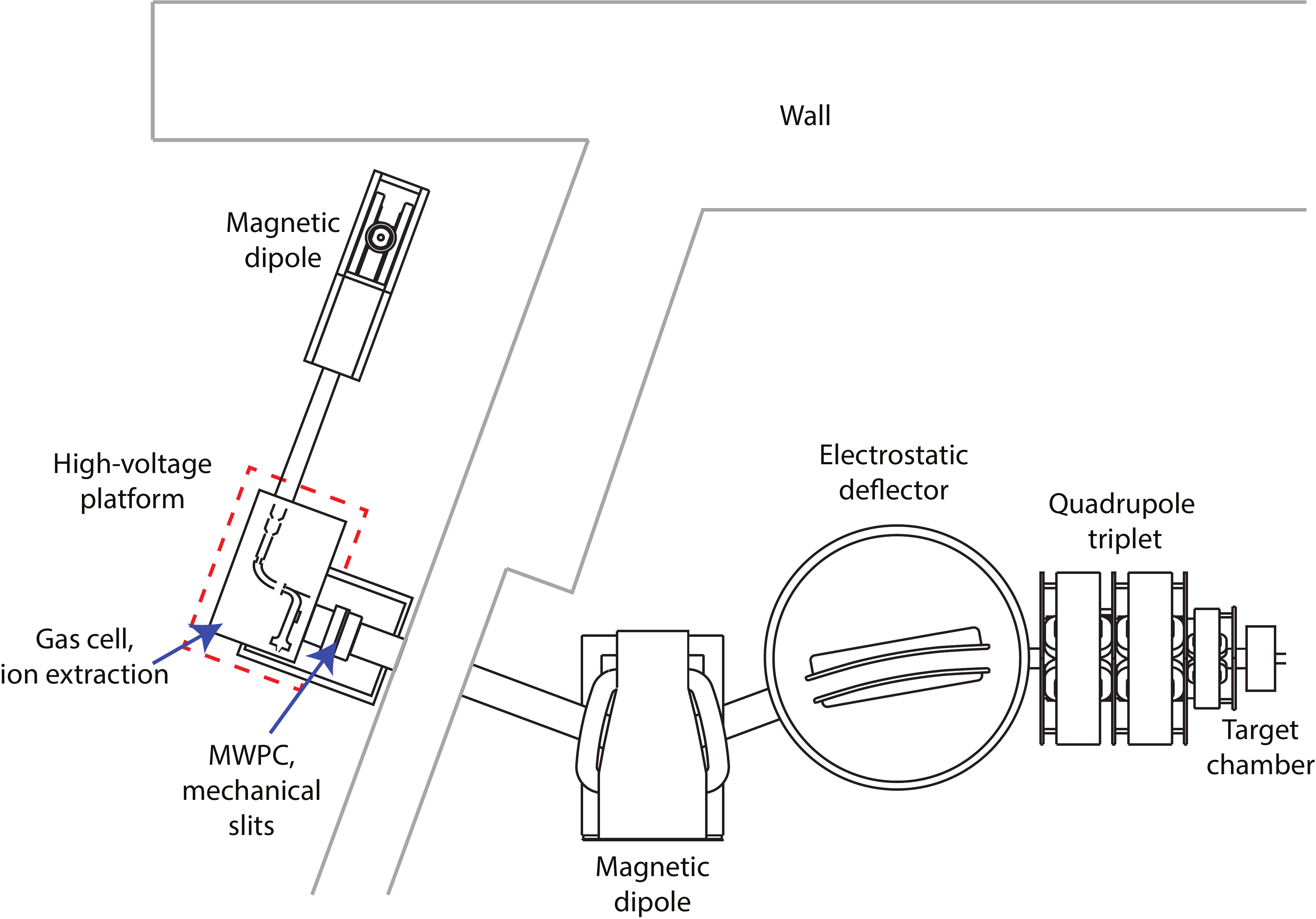}
\caption{Schematic layout of MARA and the lower level of the MARA-LEB facility. The main components of the devices are shown together with the wall separating them from the neighbouring experimental hall. The low-energy branch is not to scale}
\label{fig:layout}
\end{center}
\end{figure}

The MARA-LEB facility will be situated on two levels at the MARA separator experimental area. A preliminary layout of the facility is shown in Fig.~\ref{fig:layout}. The primary vacuum chamber, containing the gas cell, radiofrequency ion guides and extraction electrodes, will be positioned directly at the focal plane of MARA (lower level). The dipole magnet which will provide further mass separation will be on the same level and will bend the ions upwards. An electrostatic deflector will bend the ions towards the second level and direct them to the experimental stations. The second level will also host the laser room and all required power supplies.

\section{Outlook}
\label{sec:outlook}
A low-energy radioactive ion beam facility for the MARA vacuum-mode mass separator is under development. The facility will include a gas cell at the focal plane of MARA for stopping and thermalising reaction products, a laser system for resonance laser ionisation and spectroscopy, radiofrequency ion guides for transportation of the ions, a dipole magnet for mass separation and experimental setups providing different capabilities, including a compact decay station and an MR-TOF-MS for mass measurements. MARA was successfully commissioned during 2015-2016 and the first experiments using the MARA focal-plane detector system are scheduled for the summer and fall of 2016. Infrastructure funding for the MARA-LEB facility, including the laser systems, has been secured via the Academy of Finland's Research Infrastructure Funding.

\section{Acknowledgements}
The research leading to these results has received funding from the People Programme (Marie Curie Actions) of the European Union's Seventh Framework Programme (FP7/2007-2013) under REA grant agreement n$^\circ$ 626518 and the Academy of Finland under the Finnish Centre of Excellence Programme 2012-2017 (Nuclear and Accelerator Based Physics).

\end{document}